# A Novel Bistatic Joint Radar-Communication System in Multi-path Environments

Yuan Quan, Longfei Shi*, Jialei Liu and Jiazhi Ma
State Key Laboratory of Complex Electromagnetic Environment Effects on Electronics and Information System
National University of Defense Technology
Changsha, China
e-mail: qy0690@gmail.com, longfeishi@sina.com, atrluoke@sina.com, jzmanudt@163.com

*Abstract*—Radar detection and communication can be operated simultaneously in joint radar-communication (JRC) system. In this paper, we propose a bistatic JRC system which is applicable in multi-path environments. Basing on a novel joint waveform, a joint detection process is designed for both target detection and channel estimation. Meanwhile, a low-cost channel equalization method that utilizes the channel state information acquired from the detection process is proposed. The numerical results show that the symbol error rate (SER) of the proposed system is similar to that of the binary frequency shift keying system, and the signal to noise ratio requirement in multi-path environments is less than 2 dB higher compared with that in single-path environment to reach a SER of $10^{-5}$. Besides, the knowledge of the embedded information is not required for the joint detection process and the detection performance is robust to unknown information.

## I. Introduction

The increasing number of communication devices and microwave sensors brings forward the problem of spectrum congestion. Therefore, the integration of communication and radar systems has been a primary research field in recent years. The levels of integration of joint radar-communication systems fall into four categories: non-integration, coexistence, cooperation and codesign [1]. Codesign methods design the joint system from the ground up and aim to maximize the spectrum efficiency and mitigation performance [2]. This paper proposes a novel codesign method for a bistatic joint radar-communication (JRC) system in multi-path environments. The cores of the proposed system are the joint detection process which improves the spectrum efficiency and the judgement-reconstruction channel equalization process which controls the symbol error rate (SER) in multi-path environments at a low level.

In the proposed bistatic system, one of the users acts as a radar and communication transmitter, and the other acts as a radar and communication receiver. Radar and communication share the waveform, the signal processing procedure and the hardware. A scenario in vehicular sensing and communication is shown in Figure 1. The JRC waveform is transmitted by User 1, reflected by several targets and received by User 2. Consequently, User 2 can receive the transmitted information from User 1 and sense the environment without a transmitter. Note that User 2 can be a group of users, and User 1 is considered as a central node which sends information to those users and activates their sensing ability. This operation mode significantly improves the spectrum efficiency and reduces the hardware cost because several users can sense the environment without transmitters. However, two requirements are put forward to the JRC system design in this scenario. First, the detection process is operated with uncertain transmitted waveform because the embedded information is unknown to User 2. Second, a low-cost channel equalization method is necessary because the emission is transmitted through the reflection line, which leads to a lower signal to noise ratio (SNR) and serious multi-path effect.

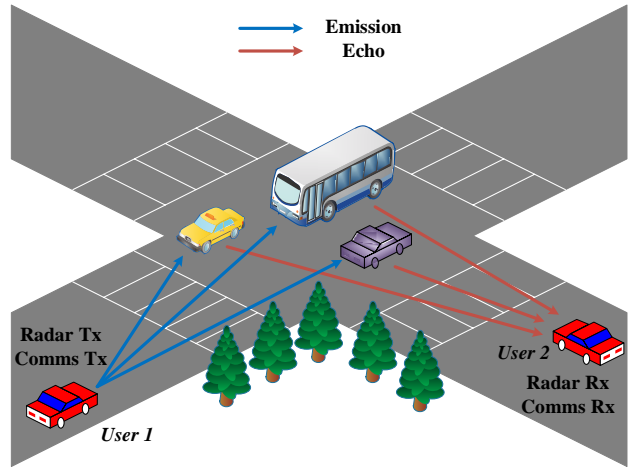

Figure 1. Bistatic JRC system in multi-path environment.

Potential solutions are found in investigated JRC systems. Information is embedded into radar waveform by the combination of linear frequency modulation (LFM) signal and minimum shift keying modulation (MSK) in [3], but the detection performance is severely affected. To reduce the influence, reduced binary phase shift keying (BPSK) is used instead of MSK [4], and the detection performance and the communication performance can be balanced by the degree of the reduced phase. However, radar waveform-based JRC system requires the full knowledge of the transmitted waveform for match filtering process, which does not satisfy the requirement of the proposed scenario. Besides, radar beam sidelobe can be utilized for simultaneous communication [5], and the anti-noise performance can be improved by sidelobe cancellation [6]. JRC system based on spatial beam design is promising. Nevertheless, we mainly focus on the main lobe at this stage. Training signal is used for channel estimation in communication system [7], and it can be modified for target detection [8]. In this way, the detection process is operated without the knowledge of the embedded information because the data signal and the training signal are separated by guard interval. Whereas, a much higher SNR is required to maintain the SER performance by channel equalization in multi-path environments. Besides, the channel equalization performance may degrade because the training signal and the data signal are separated in the time domain. Generally, JRC systems mentioned above are available in many scenarios but limited in the scenario shown in Figure 1.

Therefore, we proposed a novel JRC system to satisfy the requirements in the scenario described in Figure 1. The transmitted waveform includes two layers of modulation where the first layer embeds information into the waveform and the second layer activates the detection ability, and the communication and the detection are realized simultaneously in different layers. A joint detection process based on radar signal processing and constant false alarm rate (CFAR) detector is designed, and the channel state information (CSI) is extracted from the detection result. The channel equalization process includes a channel state judgement process and a judgement-reconstruction equalization process which make the SER in multi-path environments close to that in single-path environment.

## II. TRANSMITTED WAVEFORM AND SIGNAL PROCESSING

### A. Transmitted Waveform Model

Define $s_1$ and $s_2$ as a pair of orthogonal signals. Their cross-correlation is written as

$$R_{ij}(t) = s_i(t) \otimes s_j^*(-t), i,j=1,2 \quad (1)$$

where $\otimes$ and $(\cdot)^*$ represents the convolution operation and the conjugation operation, respectively. According to [9], the autocorrelation and the cross-correlation of ideal orthogonal signals can be described as

$$R_{11}(t) = R_{22}(t) = R(t) = \begin{cases} E_s, t=0 \\ 0, t \neq 0 \end{cases} \quad (2)$$

and

$$R_{12}(t) = R_{21}^*(-t) = 0 \quad (3)$$

where $E_s$ is the signal energy.

The transmitted JRC waveform can be expressed as

$$s(t) = \sum_{n=0}^{N-1} \left[ I_n s_1(t-nT_s) + \overline{I_n} s_2(t-nT_s) \right] e^{j\varphi_n} \quad (4)$$

where $I_n$ represents the binary number ('0' or '1') embedded in the $n$-th information symbol, $N$ is the number of the information symbols in the waveform, $T_s$ is the duration of the orthogonal signals and $\varphi_n$ is the external phase of the $n$-th information symbol. The duration of the waveform is $T=NT_s$ and the energy of $s(t)$ is $E=NE_s$.

### B. Information Demodulation

Considering a single point stationary target, the received signal can be expressed as

$$r(t) = s(t-t_0) + w(t). \quad (5)$$

The internal match filtering is conducted by matching the received signal with two match filters, and the outputs can be written as

$$r_1(t) = r(t) \otimes s_1^*(-t)$$
$$r_2(t) = r(t) \otimes s_2^*(-t). \quad (6)$$

Assuming $s_1$ and $s_2$ are ideal orthogonal, (6) can be simplified to

$$r_1(t) = \sum_{n=0}^{N-1} I_n R(t-t_0-nT_s) e^{j\varphi_n} + w_1(t)$$
$$r_2(t) = \sum_{n=0}^{N-1} \overline{I_n} R(t-t_0-nT_s) e^{j\varphi_n} + w_2(t) \quad (7)$$

where

$$w_1(t) = w(t) \otimes s_1^*(-t)$$
$$w_2(t) = w(t) \otimes s_2^*(-t). \quad (8)$$

By comparing the values of the two outputs at

$$\mathbf{t}_0 = [t_0 \ t_0+T_s \ t_0+2T_s \ ... \ t_0+(N-1)T_s]^T, \quad (9)$$

the embedded information $\mathbf{I} = [I_0...I_n...I_{N-1}]^T$ is acquired. It should be noted that $\mathbf{t}_0$ is obtained after the joint detection process, so the information demodulation process is conducted after the joint detection and the channel equalization process.

### C. Radar signal processing

The radar signal processing is conducted after the internal match filtering. The process is concluded as

$$r_{f2}(t) = r(t) \otimes h_{f1}(t) \otimes h_{f2}(t) \quad (10)$$

where

$$h_{f1}(t) = s_1^*(-t) + s_2^*(-t) \quad (11)$$

and

$$h_{f2}(t) = \sum_{n=0}^{N-1} \delta(-t-nT_s) e^{-j\varphi_n}. \quad (12)$$

(10) can be transformed into

$$r_{f2}(t) = R(t-t_0) \otimes R_2(t) + w_{f2}(t) \quad (13)$$

where

$$w_{f2}(t) = w(t) \otimes h_{f1}(t) \otimes h_{f2}(t) \quad (14)$$

and

$$R_2(t) = \sum_{n=0}^{N-1} \sum_{m=0}^{N-1} \delta[t-(n-m)T_s] e^{j(\varphi_n-\varphi_m)}. \quad (15)$$

As shown in (13) and (15), the output of the radar signal processing depends on the autocorrelations of the orthogonal signals and the external phase sequence

$$\boldsymbol{\varphi} = [\varphi_0 \ \varphi_1 ... \varphi_{N-1}]^T. \quad (16)$$

Specially, the range resolution is decided by that of the orthogonal signals, and the range sidelobes of the output can be controlled by the improvement of the orthogonality of the orthogonal signals and the optimization of the external phase sequence. Obviously, the signal processing does not require the knowledge of **I**, and the output of the radar signal processing is unrelated with **I**. Thus, the detection performance is robust to unknown embedded information.

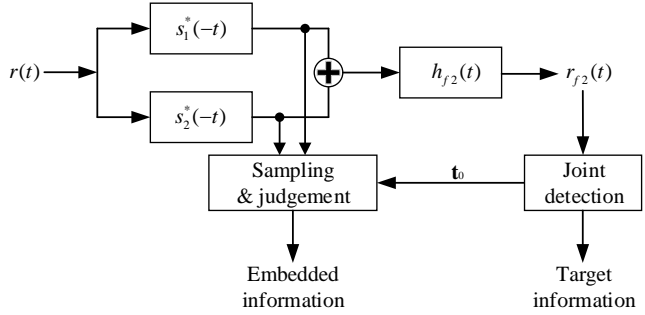

Figure 2. Information demodulation and radar signal processing in single-path environment.

The information demodulation and radar signal processing procedures are displayed in Figure 2. As mentioned in Section II-B, $\mathbf{t}_0$ is necessary for the information demodulation process. To acquire $\mathbf{t}_0$, the output of the radar signal processing is processed by CFAR detector which is widely used in radar system because of its predictable and stable detection performance [10].

### III. JOINT DETECTION AND CHANNEL EQUALIZATION

#### A. Joint Detection Process Based on CFAR Detector

In addition to the judgement time $\mathbf{t}_0$, CSI which is usually obtained by channel estimation is necessary for channel equalization in multi-path environments. In the proposed JRC system, a joint detection based on CFAR detector is designed to obtain both the target information and the CSI. However, the focus of the CSI differs from that of the target information, and the CFAR detector should be modified for communication purpose.

The detection threshold of the CFAR detector is [10]

$$T = \frac{\alpha}{M} \sum_{i=1}^{M} x_i \quad (17)$$

where $x_i$ is the $i$-th unit to be detected, $M$ is the number of units and

$$\alpha = M(P_{\text{FA}}^{-1/N} - 1). \quad (18)$$

If the noise power $\sigma^2$ is known, the detection threshold is

$$T = \sqrt{-\sigma^2 \ln P_{\text{FA}}}. \quad (19)$$

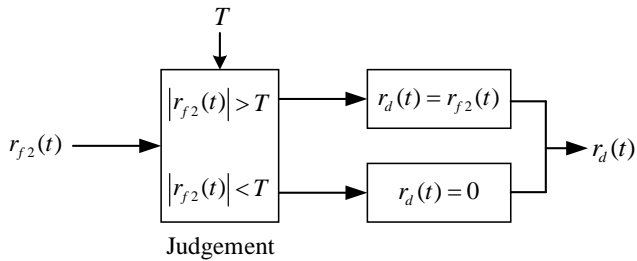

Figure 3. The joint detection process based on CFAR detector.

The joint detection process is shown in Figure 3. The detection threshold is setting according to (17) and (19). If the amplitude of $r_{f2}$ is larger than the detection threshold, both the amplitude and the phase information will be retained. Else the information will be abandoned.

The output of the joint detection process $r_d$ contains the CSI required for the judgement-reconstruction channel equalization process. The element in the channel delay sequence $\mathbf{t}_d = [t_0 \ldots t_{M-1}]^{\mathrm{T}}$ satisfies

$$\left.\frac{dr_d}{dt}\right|_{t=t_m} = 0, \left.\frac{d^2 r_d}{dt^2}\right|_{t=t_m} < 0, m = 0, 1, \ldots, M-1. \quad (20)$$

$\mathbf{t}_d$ and $r_d(\mathbf{t}_d)$ describe the CSI of the environment.

#### B. Judgement-reconstruction Channel Equalization Process

Traditional channel equalization methods such as zero-forcing and minimum mean square error methods require accurate CSI and higher SNR to reduce the SER in multi-path environments. Therefore, a novel channel equalization method based on a judgement-reconstruction process is designed for the proposed JRC system.

To improve the efficiency of the method, the channel equalization process is carried out according to the channel information matrix (**CI**) which is built based on the CSI

$$\mathbf{CI} = \begin{bmatrix} a_{0,0} & a_{1,0} & \cdots & a_{M-1,0} \\ a_{0,1} & \ddots & & \vdots \\ \vdots & & \ddots & \vdots \\ a_{0,M-1} & \cdots & \cdots & a_{M-1,M-1} \end{bmatrix} \quad (21)$$

where

$$a_{mn} = \begin{cases} 1, & \text{if } |t_m - t_n| > T_s - \tau \\ & \text{and} \\ & \text{mod}(|t_m - t_n|, T_s) \in [0, \tau) \cup (T_s - \tau, T_s] \\ 0, & \text{else} \end{cases} \quad (22)$$

in which $\tau$ is the pulse width of $r_{f2}$. The information demodulation process is shown in Figure 4. Assuming $|r_d(t_i)|$ is the maximum of $|r_d(\mathbf{t}_d)|$, if the elements in the $i$-th row of **CI** are not all zeros, the judgement-reconstruction channel equalization process will be carried out. Otherwise, the judgement process can be carried out.

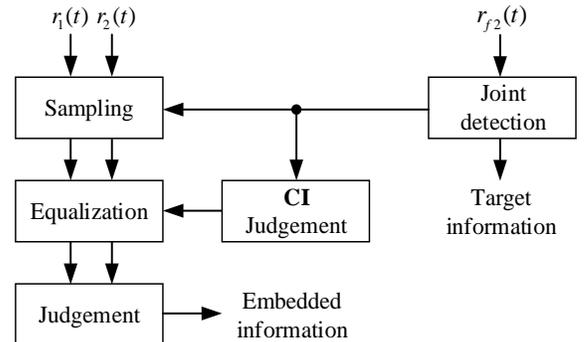

Figure 4. Information demodulation process in multi-path environments.

Assuming $t_i=t_0$ which is the delay time of the main path, the sampled signal can be described as

$$\mathbf{r}_1 = [r_{1,0}\ r_{1,1}...]^T = r_1(\mathbf{t}_0) + \sum_{k=1}^{K} r_1(\mathbf{t}_0 - T_k + t_0)$$
$$\mathbf{r}_2 = [r_{2,0}\ r_{2,1}...]^T = r_2(\mathbf{t}_0) + \sum_{k=1}^{K} r_2(\mathbf{t}_0 - T_k + t_0) \quad (23)$$

where $K$ is the number of non-zero parameters in the $i$-th row of $\mathbf{CI}$ and $T_k$ is the delay time of the $k$-th interference path. The judgement-reconstruction equalization process is expressed as

$$re_{1,0} = \begin{cases} \dfrac{r_d(t_0)\cdot e^{j\varphi_0}}{N}, & \text{if } \dfrac{r_{1,0}\cdot e^{-j\varphi_0}}{r_d(t_0)} > \dfrac{r_{2,0}\cdot e^{-j\varphi_0}}{r_d(t_0)} \\ 0, & \text{else} \end{cases}$$

$$re_{2,0} = \begin{cases} \dfrac{r_d(t_0)\cdot e^{j\varphi_0}}{N}, & \text{if } \dfrac{r_{1,0}\cdot e^{-j\varphi_0}}{r_d(t_0)} < \dfrac{r_{2,0}\cdot e^{-j\varphi_0}}{r_d(t_0)} \\ 0, & \text{else} \end{cases} \quad (24)$$

and

$$re_{1,n} = \begin{cases} \dfrac{r_d(t_0)\cdot e^{j\varphi_n}}{N}, & \text{if } \dfrac{\left[r_{1,n} - \sum_{k=1}^{K}\dfrac{r_d(T_k)}{r_d(t_0)} re_{1,n-T_k/T_s}\right]\cdot e^{-j\varphi_n}}{r_d(t_0)} \\ & > \dfrac{\left[r_{2,n} - \sum_{k=1}^{K}\dfrac{r_d(T_k)}{r_d(t_0)} re_{2,n-T_k/T_s}\right]\cdot e^{-j\varphi_n}}{r_d(t_0)} \\ 0, & \text{else} \end{cases}$$

$$re_{2,n} = \begin{cases} \dfrac{r_d(t_0)\cdot e^{j\varphi_n}}{N}, & \text{if } \dfrac{\left[r_{1,n} - \sum_{k=1}^{K}\dfrac{r_d(T_k)}{r_d(t_0)} re_{1,n-T_k/T_s}\right]\cdot e^{-j\varphi_n}}{r_d(t_0)} \\ & < \dfrac{\left[r_{2,n} - \sum_{k=1}^{K}\dfrac{r_d(T_k)}{r_d(t_0)} re_{2,n-T_k/T_s}\right]\cdot e^{-j\varphi_n}}{r_d(t_0)} \\ 0, & \text{else} \end{cases}$$

$$, 1 \le n \le N-1 \quad (25)$$

where $\mathbf{re}_1 = [re_{1,0}...re_{1,N-1}]^T$ and $\mathbf{re}_2 = [re_{2,1}...re_{2,N-1}]^T$ are the reconstructed sequences whose initial values are zero sequences.

After the judgement-reconstruction equalization process, the judgement is made on the reconstructed sequences, and the embedded information is demodulated.

## IV. SIMULATION ANALYSIS

In our simulation, the cyclic algorithm-new proposed in [11] is used to generate the orthogonal signals $s_1$ and $s_2$ with length 200.

### A. Signal Processing

The output of the signal processing for information demodulation is shown in Figure 5. 13-Barker sequence is used as the external phase sequence. By the comparison between the values of $r_1$ and $r_2$ at $\mathbf{t}_0$, the embedded information can be demodulated. Obviously, the comparison result is consistent with $\mathbf{I}$. Note that ideal orthogonality cannot be realized in real systems, and the value of $R_{12}(0)$ and $R_{21}(0)$ are not zeros in the simulation. Therefore, it is not an $E_s$-0 comparison but an $E_s$-$|R_{12}(0)|$ comparison, which makes the SER slightly higher than that of binary frequency shift keying (2FSK) system.

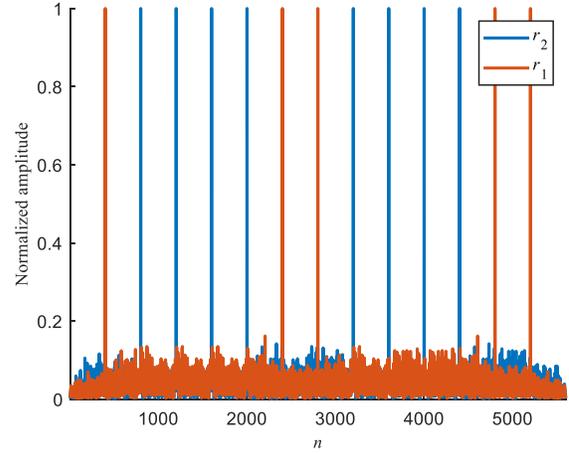

Figure 5. The comparison between $r_1$ and $r_2$ with $\mathbf{I}=[1\ 0\ 0\ 0\ 0\ 1\ 1\ 0\ 0\ 0\ 0\ 1\ 1]^T$.

The output of the radar signal processing is simulated and compared with the autocorrelations of the orthogonal signal and the external phase sequence. As shown in Figure 6, the range sidelobes of $r_{f2}$ resemble those of $R_2$, and the differences between them result from the range sidelobes of $R$. As a result, the peak sidelobe level (PSL) of $r_{f2}$ mainly depends on the larger one of the PSLs of $R_2$ and $R$. Besides, the range resolution is decided by that of the orthogonal signals because $R_2$ is a pulse sequence signal.

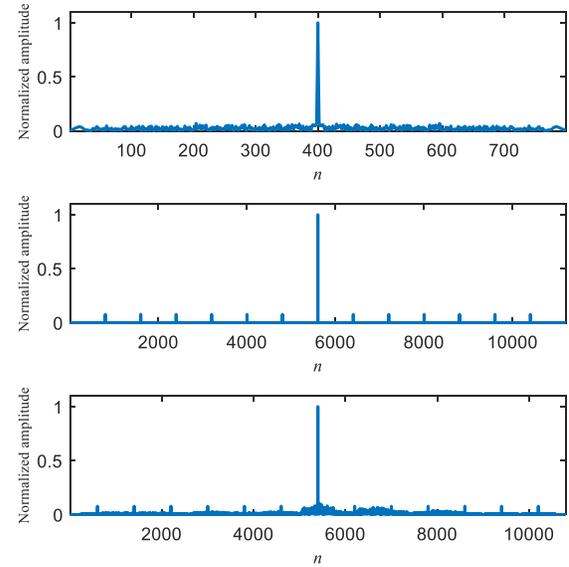

Figure 6. The autocorrelation of the orthogonal signal $R$ (first), the autocorrelation of the external phase sequence $R_2$ (second) and the output of radar signal processing $r_{f2}$ (third).

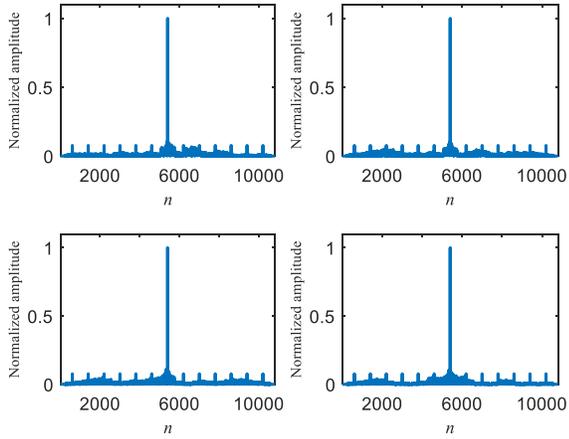

Figure 7. The output of radar signal processing $r_{f2}$ with different **I**. $\mathbf{I}_1 = [1\ 0\ 0\ 0\ 0\ 1\ 1\ 0\ 0\ 0\ 0\ 1\ 1]^T$, $\mathbf{I}_2 = [1\ 0\ 1\ 0\ 1\ 1\ 1\ 0\ 0\ 0\ 0\ 1\ 0]^T$, $\mathbf{I}_3 = [0\ 1\ 0\ 1\ 0\ 1\ 1\ 1\ 0\ 0\ 1\ 0\ 1]^T$ and $\mathbf{I}_4 = [0\ 1\ 0\ 1\ 1\ 1\ 1\ 1\ 1\ 0\ 0\ 0\ 0]^T$.

In order to prove the robustness to unknown embedded information, we compare the outputs of the radar signal processing with different embedded information in Figure 7. It can be found that there are only nuances which are caused by the non-ideal orthogonality of the orthogonal signals between these outputs, and most differences are smaller than the PSL of the outputs.

### B. Detection Probability

To show the detection performance of the proposed radar signal processing method, we analyze the detection probability ($P_d$) of the proposed method by Monte-Carlo simulations and compare it to that of the match filtering method in Figure 8. The constant false alarm probability is $10^{-6}$ and the times of the Monte-Carlo simulations are $10^8$.

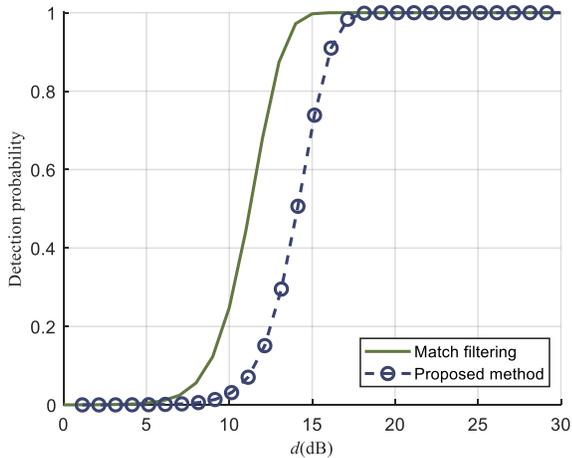

Figure 8. Detection probability of the proposed signal processing method compared with that of the match filtering method.

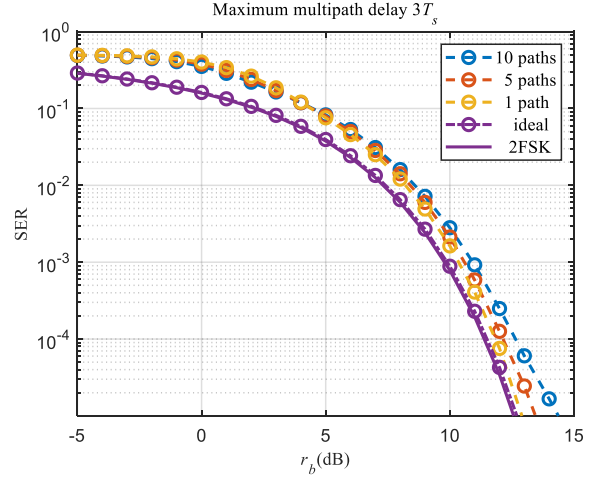

Figure 9. SER of the coherent demodulation as a function of signal to noise ratio per bit.

13-Barker sequence is used as the external phase sequence. The X-axis is $d=2E/N_0$ where $N_0/2$ is the noise power spectral density. It can be found that under a certain level of $P_d$, the required $d$ of the proposed method is 3 dB higher than that of the match filtering method, which is the cost of the robustness to unknown embedded information.

### C. Symbol Error Rate

We simulate the SER performance of the system with different numbers of paths to investigate the communication performance in multi-path environments. The times of the Monte-Carlo simulations are $6 \times 10^5$. 13-Barker sequence is used as the external phase sequence. The channel model in the simulation can be described as

$$h(t) = \delta(t-\tau_0) + \sum_{n=1}^{L-1}(a_n+b_n j)\cdot \delta(t-\tau_n) \quad (26)$$

where $L$ is the number of paths and

$$\begin{aligned}\tau_n &\sim \text{unif}(0,3T_s), \quad n=0,...,L-1 \\ a_n, b_n &\sim \text{unif}(-\sqrt{2}/2, \sqrt{2}/2), n=1,...,L-1.\end{aligned} \quad (27)$$

Demonstrated in Figure 9, the SER performance of the demodulation system closely matches with that of the coherent demodulation system of 2FSK system if ideal CSI is given in single path environment. Nuances exist because of $R_{12}(0)$ and $R_{21}(0)$.

The CSI is acquired from the joint detection process in the situation of 1 path, 5 paths and 10 paths. As shown in Figure 9, the SER performance of the system in the situation of 1 path approaches that of ideal CSI with the ascendance of SNR, which is because the CSI estimation accuracy increases with the growth of signal to noise ratio per bit ($r_b$) which equals to $2E_s/N_0/L$. In contrast, the required $r_b$ for a SER of $10^{-5}$ increases by 1 dB and 2 dB in the situation of 5 paths and 10 paths, respectively. In conclusion, the SER performance of the proposed system approaches that of the coherent demodulation system of 2FSK system in single-path environment, and the joint detection and judgement-reconstruction equalization process can maintain the SER performance of 10 paths at the same level as that of 1 path with only 2 dB higher $r_b$ requirement.

## V. CONCLUSION

In this paper, a novel JRC system in multi-path environments is proposed. The cores of the system are the joint waveform, the joint detection process and the judgement-reconstruction channel equalization process which give the system the robust detection and communication ability in multi-path environments and help to maximize the spectrum efficiency. Future work can explore the Doppler performance of the system and improve the information transmission capacity.